# QKD from a microsatellite: The SOTA experience

Alberto Carrasco-Casado*[a], Hideki Takenaka[a], Mikio Fujiwara[b],
Mitsuo Kitamura[b], Masahide Sasaki[b], Morio Toyoshima[a]
[a]National Institute of Information and Communications Technology (NICT),
Space Communications Laboratory, 4-2-1, Nukui-Kitamachi, Tokyo, Japan 184-8795
[b]National Institute of Information and Communications Technology (NICT),
Quantum ICT Advanced Development Center, 4-2-1, Nukui-Kitamachi, Tokyo, Japan 184-8795

## ABSTRACT

The transmission and reception of polarized quantum-limited signals from space is of capital interest for a variety of fundamental-physics experiments and quantum-communication protocols. Specifically, Quantum Key Distribution (QKD) deals with the problem of distributing unconditionally-secure cryptographic keys between two parties. Enabling this technology from space is a critical step for developing a truly-secure global communication network. The National Institute of Information and Communications Technology (NICT, Japan) performed the first successful measurement on the ground of a quantum-limited signal from a satellite in experiments carried out on early August in 2016. The SOTA (Small Optical TrAnsponder) lasercom terminal onboard the LEO satellite SOCRATES (Space Optical Communications Research Advanced Technology Satellite) was utilized for this purpose. Two non-orthogonally polarized signals in the ~800-nm band and modulated at 10 MHz were transmitted by SOTA and received in the single-photon regime by using a 1-m Cassegrain telescope on a ground station located in an urban area of Tokyo (Japan). In these experiments, after compensating the Doppler effect induced by the fast motion of the satellite, a QKD-enabling QBER (Quantum Bit Error Rate) below 5% was measured with estimated key rates in the order of several Kbit/s, proving the feasibility of quantum communications in a real scenario from space for the first time.

## 1. INTRODUCTION

Quantum Key Distribution (QKD) has become the most developed topic of quantum-information technologies. QKD deals with the problem of distributing cryptographic keys between two parties and its security relies on the fundamental principles of quantum physics. It is the only technology that can guarantee the unconditional security in the distribution of the keys even in the presence of an eavesdropper [1]. The key of QKD is that by evaluating the Quantum Bit Error Rate (QBER), it is possible to determine if the communication is secure or not, and if it is, the security is unconditional.

Optical fiber has been the traditional transmission channel for QKD systems, which have reached the commercial stage and have rapidly widespread in the last decade. However, reliable quantum repeaters are still a difficult challenge, hence current fiber-based QKD links are limited to distances in the order of 300 km with key rates below 1 kbit/s [2]. A truly-secure global communication network will require space-to-ground links to enable key sharing between distant locations.

Several steps towards space QKD were taken in the past before the SOTA experiment, such as using a retroreflector satellite to test the reception of polarized single-photon signals [3], using an airplane with similar characteristics to those of a space link [4], or using an air balloon with the same purpose [5]. The experiments with SOTA on August 2016 were the first ones using a quantum-limited laser source in space [6], and in 2017, China demonstrated several quantum-communication principles from space for the first time [7].

## 2. SOTA MISSION

The Japanese National Institute of Information and Communications Technology (NICT) developed a small space lasercom terminal called SOTA (Small Optical TrAnsponder) to be embarked on the microsatellite SOCRATES (Space Optical Communications Research Advanced Technology Satellite) [8]. SOCRATES was launched in May 2014 into a

600-km Low-Earth orbit (LEO), becoming the first microsatellite with high-speed lasercom capabilities, and it was operative until November 2016, more than doubling its designed lifespan. To receive the SOTA signals on the ground, an Optical Ground Station (OGS) was developed in NICT-Headquarters (Tokyo, Japan). This OGS includes a 1-m Cassegrain-type telescope (see Fig. 1, left) with a 12-m focal length, capable of tracking LEO satellites with an accuracy better than 10 arcsec.

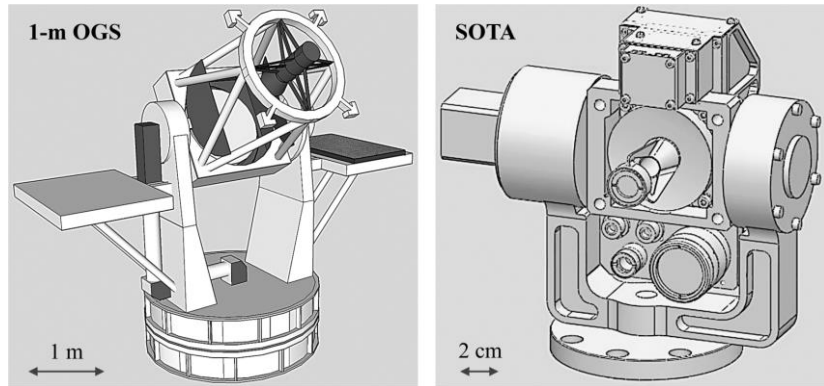

Fig. 1. 3D model of 1-m Optical Ground Station (left) and flight model of SOTA (right).

SOTA (see Fig. 1, right) was a 6-kg terminal designed to carry out several different lasercom experiments, namely, coarse/fine tracking of a ground beacon laser, 10-Mbit/s lasercom space-to-ground downlinks, transmission of the SOCRATES-camera images through lasercom, experiments with error correcting codes, interoperability with other international ground stations, and the quantum-limited experiment described in this paper. SOTA had four different laser sources to perform all these experiments: Tx1 at 976 nm with linear polarization, Tx2 and Tx3 at the ~800-nm band with non-orthogonal linear polarizations, and Tx4 at 1549 nm with circular polarization.

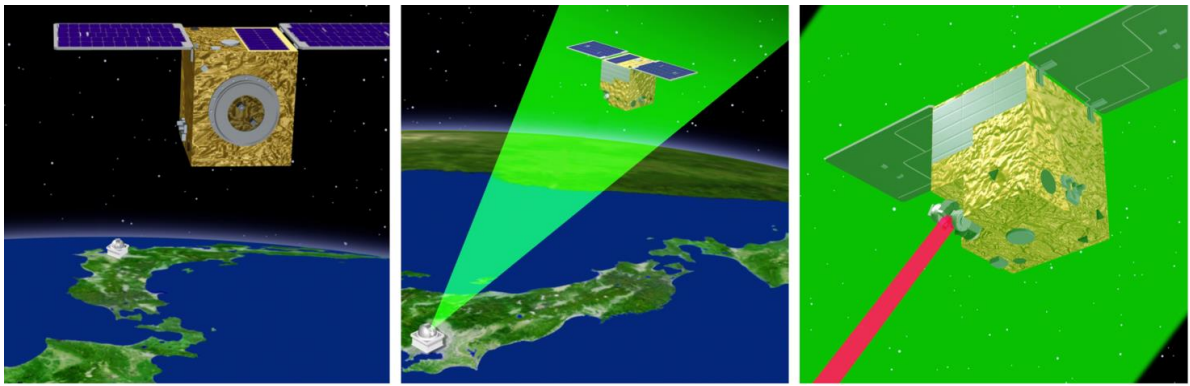

Fig. 2. SOTA approaching the NICT's OGS in Tokyo (left), the OGS transmitting a laser beacon towards SOTA (center), SOTA transmitting the lasercom signal towards the OGS (right).

The usual operation of SOCRATES on a typical SOTA pass is shown in Fig. 2. First, as the satellite approaches Japan, the OGS starts tracking by pointing the telescope towards the predicted position of SOCRATES by using its orbital information (Fig. 2, left). When the OGS elevation surpasses a certain level, it starts transmitting a powerful beacon laser with a wide divergence to cover the position uncertainty (Fig. 2, center). At a certain point, SOTA, which was blindly tracking the predicted OGS position, detects the beacon laser and starts transmitting the communication laser (Fig. 2, right).

Fig. 3 shows an example of the SOTA link budget when transmitting with the Tx4 laser at 1549 nm from a distance of 998 km at an OGS elevation of ~35°. This link budget shows a good agreement between the experimental received power and the expected one, with a difference of ~3 dB. Therefore, it is useful to validate different elements of the quantum-limited experiment, which happens at a much lower power. When comparing this link budget with the one in the quantum-limited experiment (Fig. 7), one can realize how challenging space QKD is.

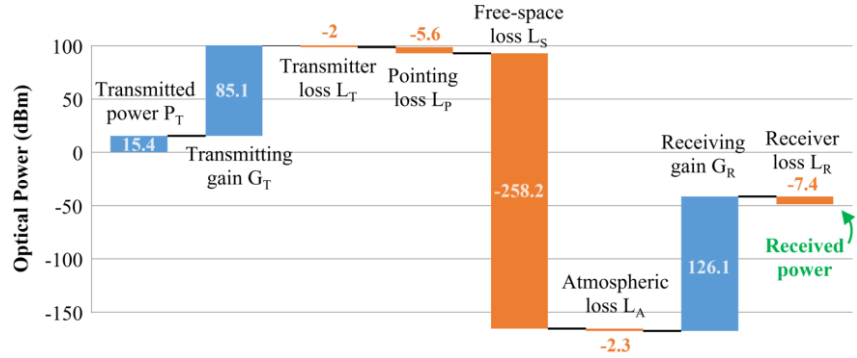

Fig. 3. Link budget for a SOTA downlink at 1550 nm at 10 Mbit/s from 998 km of distance (elevation ≈ 35°).

## 3. POLARIZATION MEASUREMENTS

BB84 is the most widely accepted QKD protocol [9], and its security has been widely studied [10], being the decoy-state its most extended version [11]. This protocol is based on the transmission of polarized single photons and its security relies on the no-cloning theorem [12]. In free-space, the information is encoded in the state of polarization since this property is well preserved in the atmosphere due to its birefringence. The effect induced by the atmospheric turbulence stays many orders of magnitude below one degree [13] [14], and the effect related to the atmospheric scattering due to aerosols and molecules in suspension is negligible, since most of the scattered photons do not reach the receiver due to the change of direction [15]. This expected behavior agrees well with the measurements carried out with SOTA, as it is shown below.

In polarization-based QKD from space, the motion of the satellite makes the reference frame change dynamically, which modifies the polarization received on the OGS in two ways, i.e. misalignments between the reference frame of the space transmitter and the ground receiver (a rotation of the polarization angle, or equivalently, a rotation of the state of polarization around the S3 axis in the Poincaré sphere), and phase delays between the orthogonal components of the state of polarization due to the different absorption of each component depending on the incident angle (making a linear polarization turn into elliptical, thus degrading its contrast ratio).

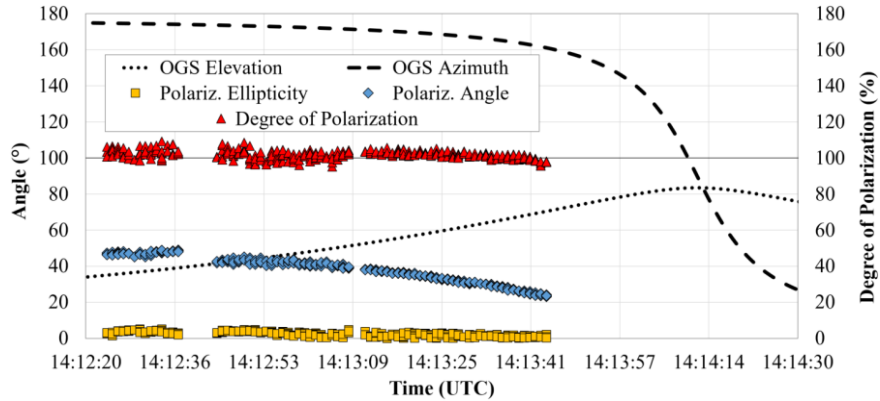

Fig. 4. Measurement of the SOTA Tx1 signal after a slant path ranging 700-1000 km on January 20th, 2016.

In 2016, several months before the quantum-limited experiments, a measurement campaign was carried out with SOTA transmitting with Tx1 and Tx4 lasers to check the polarization behavior since these lasers can be received with enough optical power to be measured by using a polarimeter in the focal plane of the OGS, as opposed to the Tx2/Tx3 lasers, which have a very-low power. Fig. 4 is an example of a real measurement showing how the polarization of the SOTA Tx1 laser was received in the NICT OGS after a slant path from LEO orbit ranging 700-1000 km. Tx1 was a linearly-polarized 976-nm laser whose signal was received with a level similar to the one shown in the Tx4 link budget, in the order of -50 dBm (Tx1 transmits with an optical power higher than Tx1 but with a wider divergence due to the lack of a fine-pointing system). The expected behavior was confirmed in these measurements with a rotation of the polarization angle as the satellite passed through its orbit, and a phase delay within the expected range (according to the simulation and experimental characterization of the telescope + receiving system).

The preservation of the degree of polarization (DOP), shown in Fig. 4 and Fig. 5, could be confirmed as well. Fig. 5 shows the histogram of the DOP of the Tx1 (left) and Tx4 (right) signals, with an average value of 100.69% and 100.32%, respectively. The deviations from 100% were satisfactorily explained by artifacts of the polarimeter, i.e. the dependence of the measurement uncertainty on the wavelength and the received power.

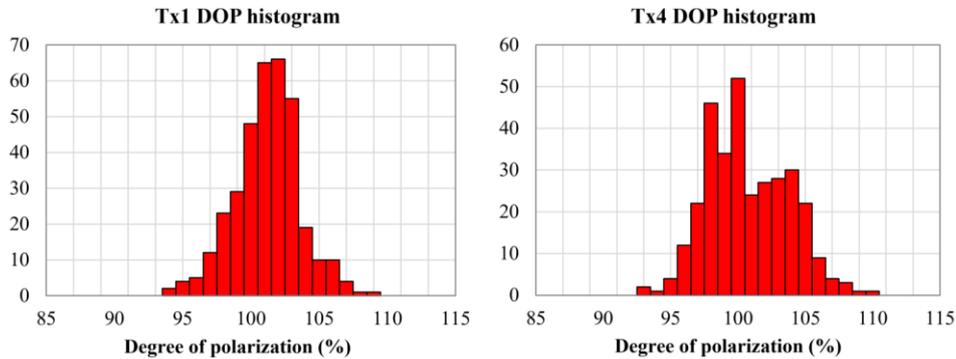

Fig. 5.. Histogram of the degree of polarization of SOTA Tx1 (left) and Tx4 (right) after LEO-to-ground propagation.

To conclude the campaign of polarization measurements, a simulation was carried out to predict the polarization angle expected to be received in the OGS as the motion of SOTA changes the reference frame between the satellite and the OGS. It is important to know beforehand how this angle changes in order to be able to apply a correction on a rotating half-wave plate in the quantum receiver to align the received reference frame with the local one. This simulation was performed by using the orbital information of SOTA in the STK (Systems Tool Kit) simulation environment, by Analytical Graphics, Inc. The simulation required to define the reference frame of the SOTA gimbal within the satellite, the reference frame of the satellite within the LEO orbit, and the reference frame of the OGS in relation to SOCRATES' orbit. Fig. 6 shows the Tx1 polarization angle received in the OGS from Tx1 (left-hand side) and Tx4 (right-hand side) during two SOTA passes. The experimental measurements showed a good agreement with the simulation.

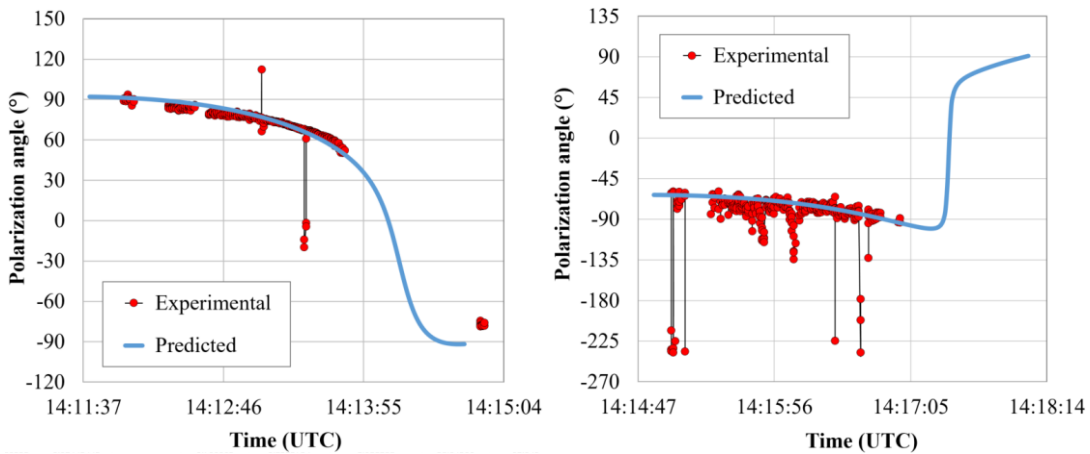

Fig. 6. Predicted an experimental polarization angle received from Tx1 (left) and Tx4 (right) during two SOTA passes.

## 4. QUANTUM-LIMITED EXPERIMENT

After the characterization of the polarization behavior of an actual laser source from space, the quantum-limited experiment campaign was carried out. Fig. 7 shows the link budget for the quantum-limited experiment using the Tx2/Tx3 lasers to compare it with the link budget of the SOTA lasercom experiments using the Tx4 laser (similar pass at a similar elevation as in Fig. 3). In terms of received power, there is a difference of more than 40 dB between both experiments, which means that the signals in the quantum-limited experiment were more than 10,000 times weaker than the ones in the conventional lasercom experiments.

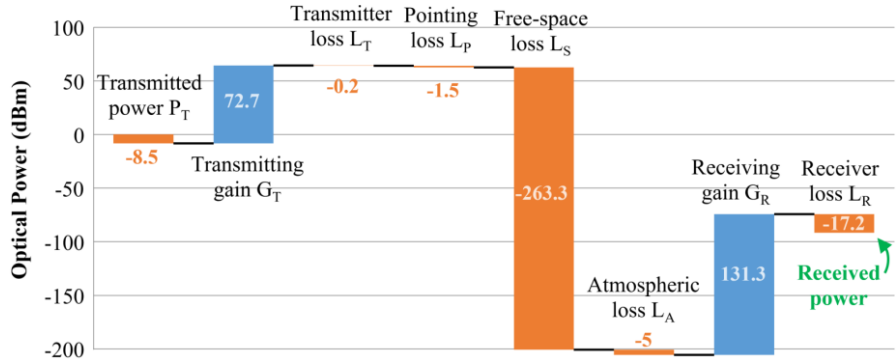

Fig. 7. Link budget for a SOTA downlink at ~800-nm band at 10 Mbit/s from 992 km of distance (elevation ≈ 35°).

The protocol implemented for the quantum-limited experiment intended to emulate the B92 QKD protocol [16]. B92 is a variation of BB84, where only one state of each polarization basis set is used, i.e. only two states instead of four. Alice encrypts the bits of the key in two linearly-polarized states at a relative angle of 45˚. Besides, since the non-orthogonal states can be unambiguously discriminated, no basis reconciliation is needed, only the time slots where Bob measures each event. As in BB84, a non-polarizing 50/50 beamsplitter models the random basis selection in Bob. This simplified QKD protocol complies with the main purpose of the experiment, which is the core of space QKD: the discrimination on the ground of non-orthogonal states at the single-photon level transmitted from a source in space.

The implementation of the transmitted signal in SOTA to establish the protocol key is shown in Fig. 8. A pseudo-random binary sequence (PRBS) was generated with a period of $2^{15}-1=32,767$, which is called PN15 (pseudo-random noise-15) at a clock frequency of 10 MHz. The bits '1' of the key were generated with the rising edges of the PN15 and the bits '0' with the falling edges. Therefore, the final key rate is half of the PN15, i.e. 5 MHz, and Tx2 and Tx3 are transmitted at half the key rate each, i.e. 2 MHz. This scheme made it possible to transmit 0°-linearly-polarized pulses through the Tx2 laser with the bits '1' of the key and 45°-linearly-polarized pulses through the Tx3 laser with the bits '0'.

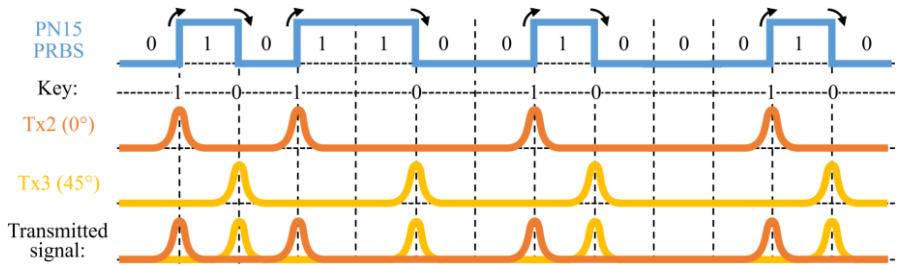

Fig. 8. Generation of the key and the transmitted signal from a PN15 PRBS in the SOTA terminal.

The optical configuration of the quantum-limited experiments carried out in August 2016 is shown in Fig. 9. In SOTA, Tx4 was transmitted to perform satellite tracking from the ground, and Tx2 and Tx3 carried the key bits as explained in the previous paragraph. Tx2 and Tx3 signals were transmitted with coarse pointing only, although Tx4 was transmitted with coarse + fine pointing, which demonstrated the feasibility of using this technology for QKD in a real application. Since fine pointing was not available for the quantum-limited signals, their divergence was widened to make up for the pointing uncertainty, and besides the pulses were brighter to compensate the big losses since the footprint on the ground was between 600 m and 1 km. Therefore, the experiment was not designed to transmit single-photon pulses, but to receive single-photon pulses. This practical implementation could be adopted in real QKD as well if the eavesdropper is assumed to be in the surroundings of the OGS instead of the surroundings of the fast-moving satellite. In any case, although this experiment was carried out in the single-photon regime on the ground, the average number of photons per bit was 0.6, implying the presence in a non-negligible number of multi-photon pulses, which is not allowed for QKD in a real application.

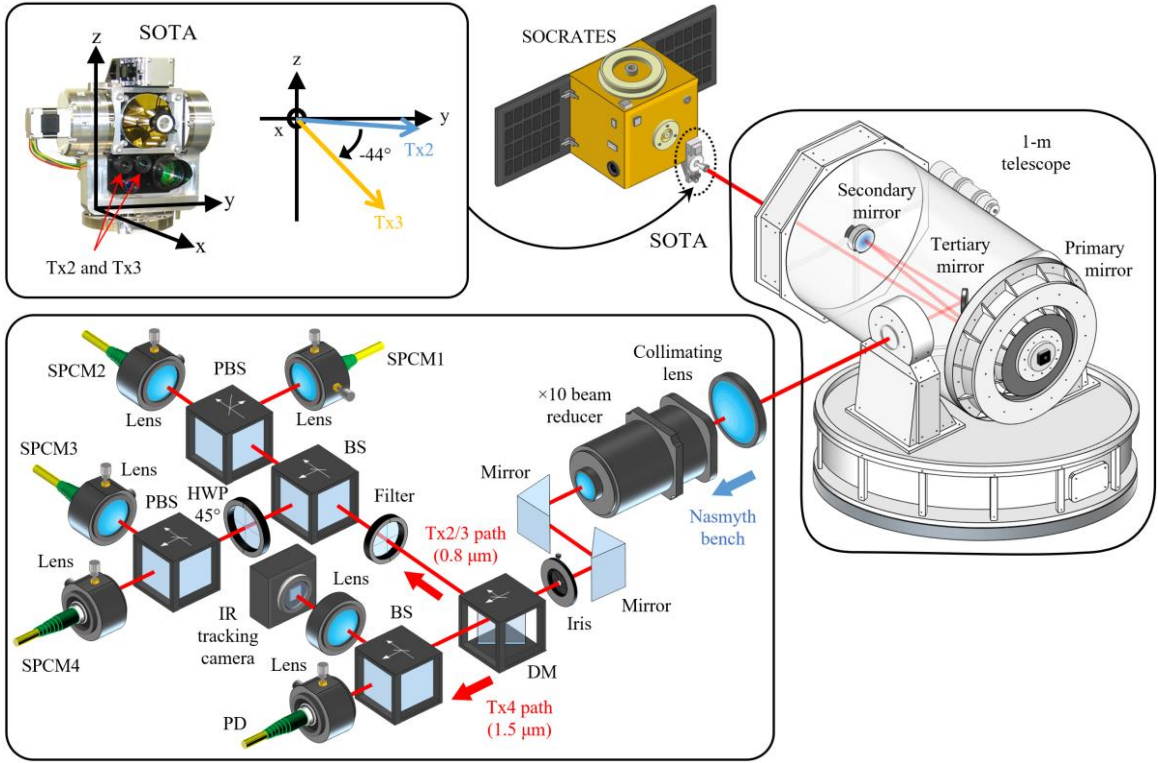

Fig. 9. Optical diagram of the SOTA quantum-limited experiment between SOCRATES and the 1-m OGS. In the quantum receiver, DM is Dichroic Mirror, BS is Beam Splitter, PBS is Polarizing Beam Splitter, HWP is Half-Wave Plate and MMF is Multi-Mode Fiber.

On the ground, the 1-m telescope gathered all the signals through the primary and secondary mirrors, and a fold tertiary mirror deflected the beam towards the Nasmyth bench, where the beam was transformed after a collimating lens and a ×10 beam reducer. Two fold mirrors used for alignment redirected the beam towards the quantum receiver, where the 1549 nm and the ~800 nm signals were separated with a dichroic mirror. The Tx4 signal was tracked with an IR camera and its power was measured. The Tx2 and Tx3 signals were detected by the conventional setup of a BB84 receiver with a beamsplitter creating two paths towards the rectilinear basis and the diagonal basis and a 45° half-wave plate before the latter. Both rectilinear and diagonal basis are identical, consisting of a polarizing beamsplitter and two focusing lenses with four multimode optical fibers in their focal plane. These optical fibers are connected to four Si-based Single-Photon Counter Modules (SPCM).

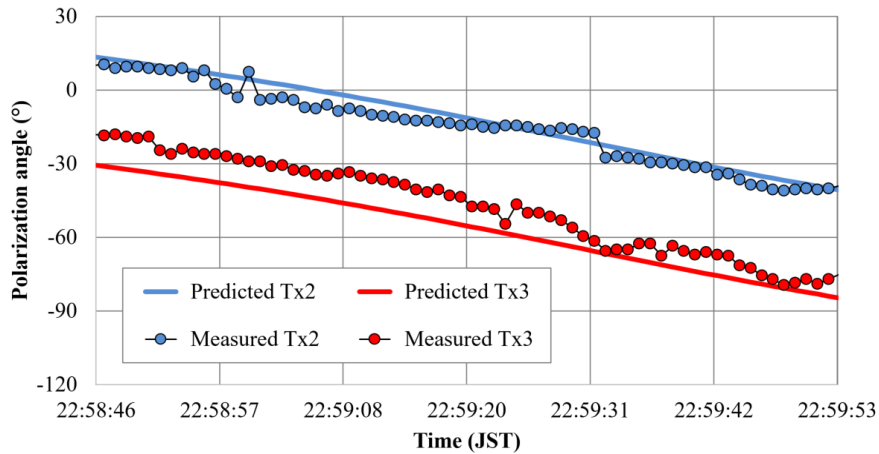

Fig. 10. Tx2 and Tx3 received polarization angle in the SOTA quantum-limited experiment.

A rotating half-wave plate was prepared for the quantum-limited experiment to compensate for the polarization reference-frame changes as the satellite moved with respect to the OGS by using the prediction shown in Fig. 6. However, in the August 2016 experiment campaign, this system could not be used (hence it is not shown in the Fig. 9 diagram). Therefore, the compensation of the reference frame was carried out in the post-processing. Furthermore, the alignment of the quantum receiver was not optimal because the same telescope was shared for different parallel experiments and the four different ports of the quantum receiver were not balanced. Additionally, their balance varied with the azimuth/elevation of the telescope, for which a characterization of the whole system was carried out observing bright stars as a reference to balance the quantum receiver in the post-processing. After all the corrections were applied to the counts detected in each port during the experiment, the received polarization angle is shown in Fig. 10 for Tx2 and Tx3, with a good agreement with the angle predicted by the SOTA orbital information.

Regarding the signal processing on the OGS, the electric SPCM signals carrying the received photon counts were time-tagged with a time-interval analyzer with a 1-ps timing resolution. Since the clock data was recovered from the SOTA signals, the Doppler shift had to be taken into account. SOCRATES was moving at ~7 km/s at a distance ranging from 700 to 1000 km, with the Doppler shift within ±200 Hz around the clock frequency (10 MHz), with a frequency drift up to ~2 Hz/s. After the synchronization with the Doppler-compensated recovered clock signal, the photon counts turned into a bit sequence that can be compared with the transmitted key from SOTA to calculate the QBER.

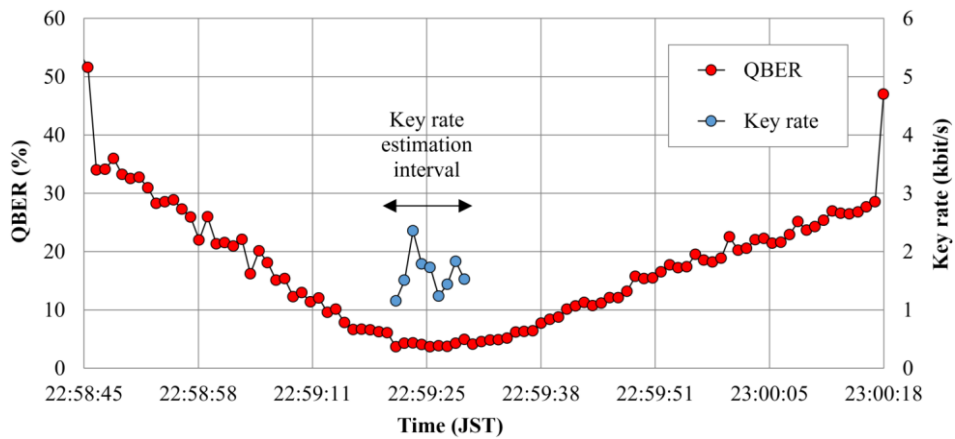

Fig. 11. Quantum Bit Error Rate and Key rate estimation in the SOTA quantum-limited experiment.

A B92 receiver has one of the following outcomes for each detected photon: either it can discriminate unambiguously between the two non-orthogonal polarization states, '1' or '0', or it is impossible to discriminate between them, i.e. there is 50% of probability of '1' and '0'. Therefore, the QBER is calculated for the unambiguous states assuming a ½ reduction in the key rate in relation to BB84 due to discarding the inconclusive counts in the sifting process. The estimated QBER for the quantum-limited experiment is shown in the Fig. 11, reaching a minimum of 3.7%, when it is assumed that the SOTA reference frame was aligned with the OGS reference frame. Before and after this, the QBER increases due to the lack of polarization tracking in the OGS. As shown in the figure, around the minimum QBER, where the polarization is aligned, the key rate was estimated to be around 1 and 2 kbit/s.

## 5. QKD SATELLITE NETWORK

The working principle behind a satellite QKD network is based on a satellite as a trusted-node linking two distant OGSs consecutively. The goal of the network is to provide the same cryptographic key to the two OGSs so that they can use it to communicate between each other either by one-time pad encryption or by a symmetric-key protocol such as AES (Advanced Encryption Standard). Fig. 12 shows a basic diagram of this working principle with its two main variants: based on QKD downlinks (left) and based on QKD uplinks (right).

In networks based on QKD downlinks, first Alice (in a satellite) transmits a Key 1 to Bob 1 (in an OGS) by establishing a QKD link, aided by the public channel 1, which is an authenticated channel based on RF or lasercom used to exchange information for key distillation, information reconciliation and privacy amplification. After some time, when the same satellite is passing over the other OGS (Bob 2), Alice transmits a Key 2, different from Key 1, by another QKD link with

Bob 2. Alice, who knows both Key 1 and Key 2, transmits Key 3, calculated as Key 3 = (Key 1) XOR (Key 2), by using the public channel 1. This way, Bob 1 will be able to calculate Bob 1 key as Key 1 = (Key 2) XOR (Key 3), and Bob 1 and Bob 2 will be able to start the secure communication encrypted by the shared Key 1 using the public channel 2, typically based on optical fiber.

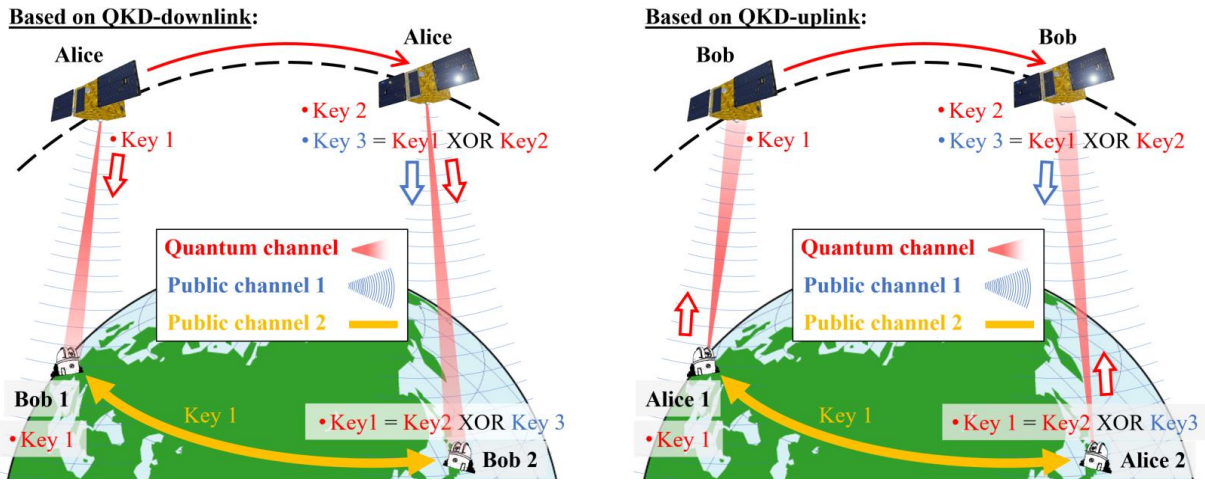

Fig. 12. Basic diagram of the working principle behind a satellite QKD network based on QKD downlinks (left) or QKD uplinks (right).

The working principle of a satellite QKD network based on QKD uplinks (Fig. 12, right) is very similar to the one based on QKD downlinks with the difference that it is Alice 1 (in an OGS) the one which starts transmitting the Key 1 to Bob (in a satellite) by a QKD link. When the satellite is passing over the other OGS (Alice 2), Alice 2 transmits the Key 2 by another QKD link with Bob, and Bob transmits the Key 3 back to Alice 2 by the public channel 1 so that Alice 2 can calculate Key 1 from Key 2 and Key 3.

The QKD fundamental principles demonstrated with SOTA can be applied to both scenarios, although there are several differences between them. First, in terms of complexity a QKD downlink allows a simpler implementation on the satellite side, the main challenges being the accurate temperature control to match the wavelengths of the four lasers of BB84, and the fine-pointing system to minimize the free-space losses. The main drawback of the QKD uplink lies in the geometrical impairment, since it is not feasible to use big apertures on the receiver side on account of being on a satellite. Conversely, in the downlink, a diffraction-limited small aperture (typically below 5 cm) can transmit a very narrow beam, which translates into a footprint below 50 meters on the ground from LEO, feasible when aided by a fine-pointing system.

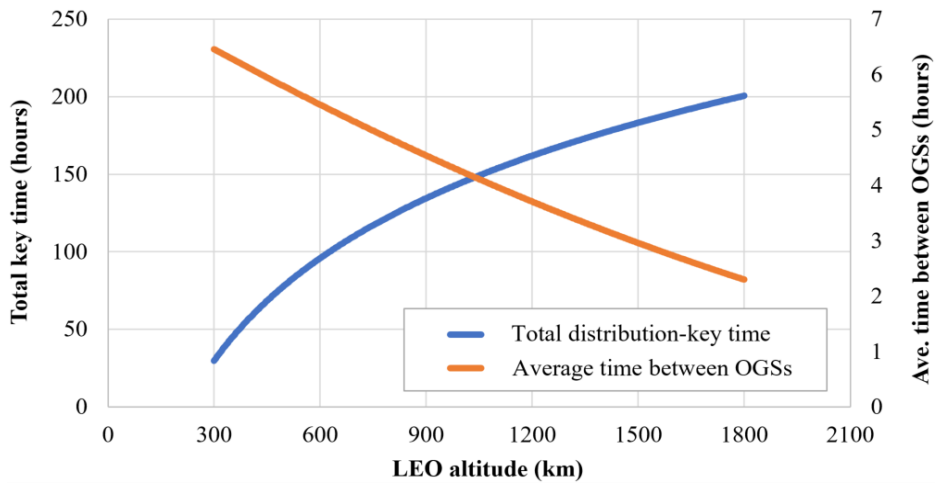

Fig. 13. Dependence of the key-transmission time and time between OGSs on the LEO altitude.

A QKD satellite network requires a different design from the one usually adopted in conventional telecom networks. The main difference comes from the working principle explained before, where it is the same satellite that must connect two OGSs consecutively without any crosslink or intermediate node. In this regard, LEO is the preferred orbit since an important parameter is the time it takes sharing the secret key between both parties. Furthermore, the launch cost to reach the orbit is lower and the radiation environment is more benign than with higher orbits. Low-Earth orbits usually have altitudes between ~200 km (below this altitude the decay is too fast due to atmospheric drag) and ~2,000 km. This allows Sun-synchronous satellites to complete multiple orbits around the Earth every day, which implies frequent key-distribution between OGSs.

Fig. 13 shows an example of how the total key-transmission time and the time between OGSs depend on the LEO altitude. This example assumed Sun-synchronous orbits during one year (2018) with a minimum OGSs elevation of 20° and the two OGSs located in Tokyo (Japan) and Madrid (Spain), which are separated by a longitude of ~135°. A successful key distribution happens each time the satellite passes over two consecutive OGSs, no matter which one starts. In each key distribution, there are two QKD sessions, one for each OGS. To calculate the key-distribution time, the shortest of these two sessions was assumed (although the difference between sessions is smaller than 2% on average). With a minimum OGS elevation of 20°, the shortest QKD session is always much longer than the time it takes to transmit a 256-bit key (the safest standard for AES), which at 1 kbps, it is 256 ms, being the average QKD session almost 2 minutes for 300-km up to over 10 minutes for 1800-km.

The total key time shown in Fig. 13 is the total key-distribution time during 2018 for each of the orbits shown in Fig. 14. The time between OGSs is the average time that takes to reach the next consecutive OGS to complete a key distribution to both OGSs. For simplicity, night-and-day links were assumed, as well as permanent cloud-free line of sight. The figure shows that the higher the orbit, the longer the total key-distribution time as well as the shorter time between OGSs. This is because OGSs observing higher orbits have a wider Field of Regard (FOR) for the same elevation than observing lower orbits, which means a longer total access time. Additionally, the wider FOR of higher orbits increases the probability that the next OGS will be available soon even though there are less Earth rotations. On the other hand, higher altitudes imply higher free-space losses, thus a lower key rate, although it could be mitigated increasing the transmitter aperture to reduce the divergence and increasing the fine-pointing accuracy.

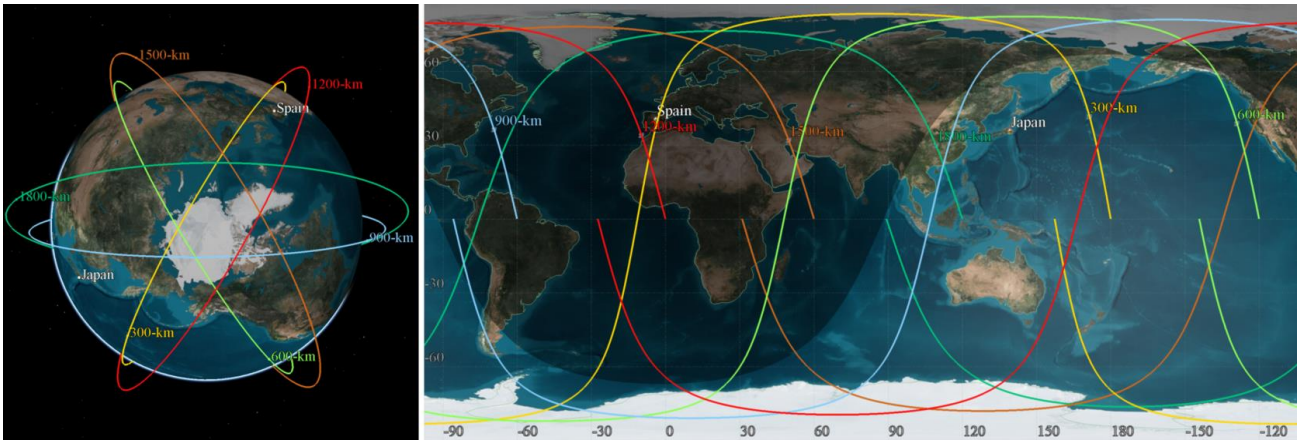

Fig. 14. Low-Earth Orbits simulated for QKD links ranging for 300 km to 1800 km during 2018 (AGI Systems Tool Kit).

## 6. CONCLUSION

Quantum communication from space, and specifically QKD, is bound to become a strategic technology to develop the future secure global communication networks. This paper summarized all the main achievements of the SOTA quantum-limited experiment, which was the first time a polarization in the 1549 nm band as well as a linear polarization has been characterized through the atmospheric propagation from a laser source in space, and the first time a quantum-limited experiment in the single-photon regime based on non-orthogonal polarization states transmitted from a space terminal has been carried out. The basic principles of a QKD network were explained and its main elements were discussed with an example of a typical QKD network linking two distant OGSs by using low-Earth orbits.

# REFERENCES

[1] N. Gisin, G. Ribordy, W. Tittel, and H. Zbinden, "Quantum cryptography," Reviews of Modern Physics, 74(1), 145–195 (2002).

[2] B. Korzh, C. C. Wen Lim, R. Houlmann, N. Gisin, M. J. Li, D. Nolan, B. Sanguinetti, R. Thew, and H. Zbinden, "Provably secure and practical quantum key distribution over 307 km of optical fibre," Nature Photonics 9, 163-168 (2015).

[3] G. Vallone, D. Bacco, D. Dequal, S. Gaiarin, V. Luceri, G. Bianco, and P. Villoresi, "Experimental satellite quantum communications," Physical Review Letters, 115(4), 040502 (2015).

[4] S. Nauerth, F. Moll, M. Rau, C. Fuchs, J. Horwath, S. Frick, and H. Weinfurter, "Air-to-ground quantum communication," Nature Photonics 7(5), 382–386 (2013).

[5] J. Wang, B. Yang, S. Liao, L. Zhang, Q. Shen, X. Hu, J. Wu, S. Yang, H. Jiang, Y. Tang, B. Zhong, H. Liang, W. Liu, Y. Hu, Y. Huang, B. Qi, J. Ren, G. Pan, J. Yin, J. Jia, Y. Chen, K. Chen, C. Peng, and J. Pan, "Direct and full-scale experimental verifications towards ground-satellite quantum key distribution," Nature Photonics 7(5), 387–393 (2013).

[6] H. Takenaka, A. Carrasco-Casado, M. Fujiwara, M. Kitamura, M. Sasaki, and M. Toyoshima, "Satellite-to-ground quantum-limited communication using a 50-kg-class microsatellite," Nature Photonics, 11, 502-508 (2017).

[7] J. Yin, Y. Cao, Y. Li, S. Liao, L. Zhang, J. Ren, W. Cai, W. Liu, B. Li, H. Dai, G. Li, Q. Lu, Y. Gong, Y. Xu, S. Li, F. Li, Y. Yin, Z. Jiang, M. Li, J. Jia, G. Ren, D. He, Y. Zhou, X. Zhang, N. Wang, X. Chang, Z. Zhu, N. Liu, Y. Chen, C. Lu, R. Shu, C. Peng, J. Wang, and J. Pan, "Satellite-based entanglement distribution over 1200 kilometers," Science, 16, 1140-1144 (2017).

[8] A. Carrasco-Casado, H. Takenaka, D. Kolev, Y. Munemasa, H. Kunimori, K. Suzuki, T. Fuse, T. Kubo-Oka, M. Akioka, Y. Koyama, and M. Toyoshima, "LEO-to-ground optical communications using SOTA (Small Optical TrAnsponder) – Payload verification results and experiments on space quantum communications," Acta Astronautica, 139, 377-384 (2017).

[9] C. H. Bennett, and G. Brassard, "Quantum cryptography: public key distribution and coin tossing," in Proc. Of IEEE Int. Conf. on Computers, Systems and Signal Processing (IEEE, 1984), pp. 175–179.

[10] P. W. Shor, and J. Preskill, "Simple Proof of Security of the BB84 Quantum Key Distribution Protocol," Physical Review Letters, 85, 441-444 (2000).

[11] M. Lucamarini, J. F. Dynes, B. Fröhlich, Z. Yuan, and A. J. Shields, "Security bounds for efficient decoy-state quantum key distribution," IEEE Journal of Selected Topics in Quantum Electronics, 21(3), 6601408 (2015).

[12] W. K. Wootters, and W. H. Zurek, "A single quantum cannot be cloned," Nature, 299, 802-803 (1982).

[13] D. H. Höhn, "Depolarization of a laser beam at 6328 Å due to atmospheric transmission," Applied Optics, 8(2), 367-369 (1969).

[14] B. Crosignani, P. Di Porto, and Steven F. Clifford, "Coupled-mode theory approach to depolarization associated with propagation in turbulent media," Applied Optics, 27(11), 2183-2186 (1988).

[15] M. Li, P. Lu, Z. Yu, L. Yan, Z. Chen, C. Yang, and X. Luo, "Vector Monte Carlo simulations on atmospheric scattering of polarization qubits," Journal of the Optical Society of America A, 30(3), 448-454 (2013).

[16] C. H. Bennett, "Quantum cryptography using any two nonorthogonal states," Physical Review Letters, 68, 3121-3124 (1992).